\documentclass[twocolumn,showpacs,amsmath,amssymb,nofootinbib]{revtex4-1}

\newtheorem{thm}{Theorem}[section]

\newtheorem{prop}[thm]{Proposition}

\def\beq{\begin{eqnarray}}
\def\eeq{\end{eqnarray}}

\begin{document}

\title{Curvature invariant characterization of event horizons \\ of
    four-dimensional black holes conformal to stationary black holes}

\author{David D. McNutt}
\email{ david.d.mcnutt@uis.no}
\affiliation{Faculty of Science and Technology,\\ 
                         University of Stavanger, 
                         N-4036 Stavanger, Norway         }
%
%
\date{\today} 

\begin{abstract}

We introduce three approaches to generate curvature invariants that transform covariantly under a conformal transformation of a four dimensional spacetime. For any black hole conformally related to a stationary black hole, we show how a set of conformally covariant invariants can be combined to produce a conformally covariant invariant that detects the event horizon of the conformally related black hole. As an application we consider the rotating dynamical black holes conformally related to the Kerr-NUT-(Anti)-de Sitter spacetimes and construct an invariant that detects the conformal Killing horizon along with a second invariant that detects the conformal stationary limit surface. 

In addition, we present necessary conditions for a dynamical black hole to be conformally related to a stationary black hole and apply these conditions to the ingoing Kerr-Vaidya and Vaidya black hole solutions to determine if they are conformally related to stationary black holes for particular choices of the mass function. While two of the three approaches cannot be generalized to higher dimensions, we discuss the existence of a conformally covariant invariant that will detect the event horizon for any higher dimensional black hole conformally related to a stationary black hole which admits at least two conformally covariant invariants, including all vacuum spacetimes.




%
\end{abstract}

\maketitle

It is possible to generate solutions of the Einstein equation by applying a conformal transformation to a known solution and solving the resulting differential equations for a particular conformal factor (see references in \cite{Stephani:2003tm,McNuttPage}). In particular, dynamical black hole solutions have been produced using a conformal transformation from a stationary black hole solution. An example of such a solution is the Sultana-Dyer metric:
\begin{eqnarray}
 ds^2 &=& t^4 \biggl[ - \left( 1 - \frac{2m}{r} \right) dt^2 + \frac{4m}{r} dt dr \\
& & +  \left( 1 - \frac{2m}{r} \right) dr^2 + r^2 (d\theta^2 + \sin^2 \theta d\phi^2)\biggl], \nonumber
\end{eqnarray}
\noindent which is conformally related to the Schwarzschild solution and models an expanding black hole in the asymptotic background of the Einstein-de Sitter universe with a two-fluid matter source consisting of a null fluid and pure dust \cite{SD2005}. The event horizon of the Sultana-Dyer metric will be the image of the Killing horizon of the Schwarzschild black hole under a conformal transformation, and hence is a conformal Killing horizon.

The Sultana-Dyer metric belongs to the non-rotating (a=0) subclass of the Thakurta metrics \cite{Thakurta}, which are generated from a conformal transformation of the Kerr solution with a conformal factor dependent on the Boyer-Lindquist time coordinate: 
\begin{eqnarray}
 ds^2 &=& \alpha^2(t) \biggl[ - \left( 1 - \frac{2Mr}{\Sigma} \right) dt^2 + \frac{\Sigma}{\Delta} dr^2 + \Sigma d\theta^2 \\ 
 && + \left[ (r^2+a^2)+ \frac{2Mra^2 \sin^2 \theta}{\Sigma}\right] \sin^2 \theta d\phi^2. \nonumber \\ 
 && - \frac{4Mra\sin^2 \theta}{\Sigma} d\phi dt \biggl], \nonumber   
\end{eqnarray}
\begin{eqnarray}
& \Sigma = r^2 +a^2 \cos^2 \theta,~~\Delta = r^2 +a^2-2Mr. & \nonumber  
\end{eqnarray} 
\noindent Motivated by the Thakurta solution we will employ spacetimes conformally related to the Kerr-NUT-(Anti)-de Sitter solution \cite{PlebDem1976,Griffiths2005,Griffiths2007} as an illustrative example. We emphasize that the proposed conformally covariant horizon detecting invariants presented in this paper will apply to any four dimensional (4D) black hole conformally related to a stationary black hole.  

In general, locating the event horizon of a black hole is  difficult, as the horizon depends upon the future evolution of the spacetime.  However, for particular classes of black hole spacetimes the event horizon can be identified locally. For example, if the spacetime is stationary then knowledge of the hypersurface metric and extrinsic curvature at any moment in time will be sufficient to determine the entire spacetime and locate the horizon. 

For a stationary spacetime, if we know the Killing vector field which acts as the null generator on the event horizon then the horizon can be located by the vanishing of the squared norm of this Killing vector field. If the related Killing vector field is not known, we can determine the local comohogeneity, $n$, of the spacetime and compute the squared norm of the wedge products of $n$ linearly independent gradients of scalar polynomial curvature invariants (SPIs), which will vanish on the stationary horizon \cite{AbdLak2015,Page:2015aia}.

For non-stationary spacetimes, there is no general procedure to generate invariants that detect the event horizon.  However, as the location of the event horizon is a conformal invariant, the horizon can be found for any black hole metric that is conformal to a stationary metric, assuming the conformal factor is known \cite{Page:2015aia}. If the conformal factor is {\it not} known, this approach will no longer locate the event horizon of the conformally related black hole.

It is reasonable to assume that a SPI will be able to detect the event horizon of a black hole metric that is conformal to a stationary metric. Such a SPI must not only vanish on the horizon of the stationary metric, but also remain zero on the horizon under a conformal transformation of the metric. To be precise, we would like an invariant that vanishes on a stationary horizon and under a conformal transformation the invariant transforms as a power of the conformal factor with no terms involving derivatives of the conformal factor (which could be nonzero on the event horizon).

To construct a conformally covariant invariant that will detect the horizon, we introduce three approaches to generate conformally covariant invariants and construct invariants that are unaffected by conformal transformations in 4D. As an application we show that the conformal Killing horizon\footnote{This surface corresponds to the event horizon when the conformal factor goes to a constant at null infinity.} of a dynamical black hole conformally related to the 4D Kerr-NUT-(Anti)-de Sitter solution is detected by the vanishing of a conformally covariant invariant and that the conformal stationary limit surface \cite{SD2004} is detected by a conformally covariant invariant as well.

Assuming conformally covariant invariants can be constructed in any dimension, we establish necessary conditions to determine if a dynamical black hole is conformally related to a stationary black hole. We apply these conditions to the ingoing Kerr-Vaidya and Vaidya black holes to show when the choice of a particular mass function allows for a conformal transformation to a stationary black hole. Unlike the 4D case, in higher dimensions the existence of conformally covariant SPIs is no longer guaranteed. In the class of spacetimes admitting two conformally covariant SPIs, we show it is possible to produce additional conformally covariant invariants and construct a conformally covariant invariant that detects the horizon of any dynamical black hole conformally related to the $D$-dimensional Kerr-NUT-Anti-de Sitter metric.

\begin{section}{Construction of Conformally Covariant Invariants}

The invariants which are known to detect the stationary horizon \cite{AbdLak2015,Page:2015aia} will pick up derivative terms when a generic conformal transformation is applied to the metric, implying that they may not remain zero on the event horizon of a black hole spacetime that is conformal to a stationary metric.  In this paper we consider the problem of constructing a SPI that vanishes on the horizon of any black hole metric that is conformal to a 4D stationary spacetime, with the conformal factor being a smooth function over the spacetime which does not vanish at the event horizon. 

As this construction will rely on the approach given in \cite{Page:2015aia}, we require at most three functionally independent conformally covariant SPIs. To produce these invariants we will use the following SPIs obtained from the Weyl tensor and its first covariant derivative:
\begin{eqnarray}
& I_1 \equiv C_{abcd}\;C^{abcd},&~~ 
 \label{defi1} \;\\
& I_2 \equiv C{^*}_{abcd}\;C^{abcd},&~~ 
 \label{defi2} \;\\
& I_3 \equiv C_{abcd;e}\;C^{abcd;e},&
 \label{defi3}\;\\
& I_{3a} \equiv C_{abcd;e}\;C^{ebcd;a},&
 \label{defi3a}\;\\
& I_4 \equiv C{^*}_{abcd;e}\;C^{abcd;e},&
 \label{defi4}\;\\
& I_{4a} \equiv C{^*}_{abcd;e}\;C^{ebcd;a},&
 \label{defi4a}\;\\
& I_5 \equiv I_{1;e}I_1^{\ ;e},&
 \label{defi5}\;\\
& I_6 \equiv I_{2;e}I_2^{\ ;e}.&
 \label{defi6}\;
\end{eqnarray}
\noindent where $ {C}_{abcd} $ is the Weyl tensor and $ {C^{*}}_{abcd} $ is its Hodge dual. Applying a conformal transformation of the metric, the transformation rules for $I_3$, $I_{3a}$, $I_4$, $I_{4a}$, $I_5$ and $I_6$ pick up first derivatives of the conformal factor.

The terms involving the derivatives of the conformal factor can be nonzero on the event horizon of a generic metric conformal to a stationary black hole metric. However, by combining the invariants \eqref{defi1}-\eqref{defi6} an invariant can be  constructed that transforms by being multiplied by a negative power of the conformal factor under a conformal transformation of the metric  \cite{McNuttPage}: 
\begin{equation}
 J_4 \equiv 6 I_1 I_3 -16 I_1 I_{3a} +  I_5.
 \label{J}
\end{equation}
\noindent While $J_4$ was originally intended for spacetimes conformal to static spherically symmetric metrics, as long as the conformal factor does not vanish on the event horizon, $J_4$ is zero on the event horizon for any smooth 4D metric conformal to a stationary black hole of cohomogeneity $n=1$. 

To see this, consider a generic nonvacuum metric $g_{ab}$ which is not conformal to a static spherically symmetric metric or a Ricci flat metric and a conformally transformed metric $\hat{g}_{ab} = e^{2U}g_{ab}$, then the invariant $\hat{J}_4$ for the metric $\hat{g}_{ab}$ is related to the invariant $J_4$ for the metric $g_{ab}$ by
\begin{equation}
 \hat{J}_4 = e^{-10U}(J_4 + A^a U_{;a})
 \label{Jtransform}\;,
\end{equation}
where
\begin{equation}
A_a = -4 I_{1,a} - 16 W^e_{~ae} + 16 W_{a~e}^{~e} + 64 \bar{W}^e_{~ea} 
\label{A}\;,
\end{equation}
\noindent and 
\begin{eqnarray}
& W^a_{~bc} = C_{bdef;c}C^{adef} & \label{Weqn}, \\
& \bar{W}^a_{~bc} = C_{bdce;f}C^{fdae}. & \label{Wbeqn}
\end{eqnarray}

\noindent Using the Newman-Penrose (NP) formalism \cite{Stephani:2003tm} to express the Weyl tensor and the covariant derivative of the Weyl tensor in terms of the NP curvature scalars $\Psi_{i},~i \in [0,4]$, their frame derivatives and the spin coefficients, it can be shown that $A_a$ will vanish for all 4D spacetimes \cite{McNuttPage}. 

The vanishing of $A_a$ in 4D is due to the trace free condition for the Weyl tensor and the algebraic Bianchi identity. This implies that any conformally covariant tensor sharing the indicial symmetries of the Weyl tensor will yield a first order invariant that will be conformally covariant. We will call a tensor {\it Weyl-like} if it shares the indicial symmetries of the Weyl tensor. In four dimensions, the Hodge dual of the Weyl tensor is conformally covariant and Weyl-like, and so we can compute a similar invariant to $J_4$:

\begin{eqnarray}
K_4 \equiv 6 I_2 I_4 - 16 I_2 I_{4a} + I_6, \label{K4}
\end{eqnarray} 

\noindent which will transform in a covariant manner under a conformal transformation for any 4D spacetime:

\begin{eqnarray} \hat{K}_4 = e^{-10U} K_4.\label{K4t} \end{eqnarray}

Assuming $I_1 \neq I_2$, $I_2 \neq 0$, and $I_1, J_4$ and $K_4$ are functionally independent, we can construct the following rational invariants which are {\em conformal invariants}, in the sense that they are invariant under conformal transformations: 

\begin{eqnarray} I_1' \equiv \frac{I_1}{I_2},~ J_4' = \frac{J_4}{I_2^{\frac52}},~ K_4' \equiv \frac{K_4}{I_2^{\frac52}}. \label{CIInvs} \end{eqnarray}

\noindent The conformal invariants $I_1',~J_4'$ and $K_4'$ will be functionally independent, and so we can compute the norm of the wedge product of the exterior derivatives of these invariants to produce an invariant that will detect the horizon for any black hole conformal to a stationary black hole \cite{Page:2015aia}.

In the case that $I_2$ vanishes or one of the invariants is functionally dependent on the others, an additional invariant of higher degree can be constructed by taking the square of the Weyl tensor,

\begin{eqnarray}
C^2_{abcd} = \frac{1}{\sqrt{I_1}} C_{ab}^{~~ef}C_{efcd},
\end{eqnarray}

\noindent and then subtracting  $C^2_{a[bcd]}$ and the appropriate tensor products of the trace two-tensor with the metric to produce a new conformally covariant tensor $W_{abcd}$ satisfying $W^a_{~bac} = 0$ and $W_{a[bcd]} =0$. Since $W_{abcd}$ is conformally covariant and Weyl-like, we can compute the following invariants:  
\begin{eqnarray}
{^W}I_1 &\equiv & W_{abcd}W^{abcd}, \\
{^W}I_3 &\equiv & W_{abcd;e}W^{abcd;e}, \\
{^W}I_{3a} &\equiv & W_{abcd;e}W^{ebcd;a}, \\
{^W}I^5 &\equiv & {^W}I_{1;a}{^W}I^{;a}.
\end{eqnarray}
\noindent These can be combined to produce an invariant,
\begin{eqnarray}
{^W}J_4 \equiv  6 {^W}I_1 {^W}I_{3a} -16{^W}I_1 {^W}I_3 + {^W}I_5.
\end{eqnarray} 

\noindent  The invariant ${^W}J_4$ transforms under a conformal transformation in the following manner:

\begin{eqnarray}
{^W}\hat{J}_4 = e^{-10U} [{^W}J_4].
\end{eqnarray}

\noindent We note that ${^W}\hat{J}_4$ is a SPI of higher degree than $J_4$ and will not, in general, be divisible by $J_4$. Of course, in 4D this procedure can be applied to any conformally covariant Weyl-like tensor to generate new conformally covariant invariants of higher degree. We may also combine any two conformally covariant Weyl-like tensors to produce SPIs similar to $I_2$, $I_4$, $I_{4a}$, $I_6$ and $K_4$.

For a black hole spacetime of cohomogeneity $n\leq 3$, the first order invariants may not provide a full set of functionally independent invariants, and so we must continue to higher order derivatives to acquire $n$ functionally independent invariants. In particular, for type {\bf D} spacetimes we may have to compute up to the fifth covariant derivative of the Weyl tensor to acquire $n$ functionally independent invariants \cite{Stephani:2003tm,Collins90,Collins91}. Taking the square of the norm of the gradient of any curvature invariant which is unchanged by a conformal transformation produces a new conformally covariant invariant. 

For example, given two conformal invariants $\hat{I} = I$ and $\hat{J} = J$ of order $p$ and $p'$ respectively, then we may produce the following conformally covariant invariants of higher order:
\begin{eqnarray}
& |\nabla I|^2 \equiv  I_{;a}I^{;a},&  \nonumber \\
&|\nabla J|^2 \equiv J_{;a}J^{;a},& \\
&\nabla I . \nabla J  \equiv I_{;a}J^{;a}.& \nonumber
\end{eqnarray}  
\noindent Dividing by  appropriate powers of $\sqrt{I_1}$ yields conformal invariants of higher order. The differential Bianchi identities and the Ricci identities ensure that new functionally independent invariants are introduced by expressing the gradients of the invariants in terms of frame derivatives. 

\end{section}

\begin{section}{Rotating Black Holes Conformally Related to the Kerr-NUT-(Anti)-de Sitter Spacetime in Four Dimensions}

We will employ the NP formalism to work with three conformally covariant SPIs in order to construct a simple conformally covariant SPI that will detect the conformal Killing horizon. For a black hole conformally related to a stationary spacetime, this surface is defined as a null surface for which a conformal Killing vector field, corresponding to a Killing vector field in the stationary black hole spacetime, has vanishing norm and does not identically vanish. If the conformal factor goes to a constant at null infinity, the conformal Killing horizon will correspond to the event horizon of the black hole \cite{SD2004}. 

The horizon Killing vector field, ${\bf K}$, in the original stationary black hole spacetime either coincides with the stationary Killing vector field, or the spacetime admits at least one axial Killing vector field ${\bf M}$ for which $[{\bf K}, {\bf M}] = 0$. This second case implies that the black hole is asymptotically flat, rotating and that the horizon Killing vector field, ${\bf \tilde{K}}$ is a linear combination of the Killing vector fields ${\bf K}$ and ${\bf M}$. For stationary rotating black holes there is an additional non null surface on which the norm of a nonvanishing Killing vector field, ${\bf K}$, is zero, called the stationary limit surface. Under a conformal transformation this surface is mapped to a conformal stationary limit surface which is defined as a surface on which a nonvanishing conformal Killing vector field ${\bf \hat{K}}$ has vanishing norm.

 As the horizon detecting invariant relies on the existence of a Killing horizon \cite{Page:2015aia}, it cannot be guaranteed to detect the stationary limit surface (also known as a {\it ergosurface}) of the original Kerr-NUT-(Anti)-de Sitter solution since it is not a Killing horizon. While the invariant $Q_1$ will detect the stationary limit surface \cite{AbdLak2015} for the Kerr-NUT-(Anti)-de Sitter solution \cite{GANG}, we note that this invariant will not transform covariantly under a conformal transformation and hence will not detect the conformal stationary limit surface \cite{SD2004}. We introduce a new invariant, $J_{{\it ergo}}$ that transforms covariantly under a conformal transformation and vanishes on the conformal stationary limit of any rotating black hole conformally related to the Kerr-NUT-(Anti)-de Sitter solution.

We consider a conformal transformation applied to the 4D Kerr-NUT-(Anti)-de Sitter metric defined in \cite{GANG}, with arbitrary conformal factor $U(t,r,\theta,\phi)$: 

\beq d\hat{s}^2 &=& e^{2U} \bigg( -\frac{Q}{R^2} \left[ dt - \left(a\sin^2 \theta + 4 l \sin^2 \frac{\theta}{2}\right) d\phi \right]^2  \nonumber \\ &&+ \frac{P}{R^2}\left[ a dt - \left(r^2 +(a+l)^2\right) d\phi \right]^2 \label{KNNAmetric} \\ && + \frac{R^2}{Q}dr^2 + \frac{R^2}{P} \sin^2 \theta d\theta^2 \bigg),
\nonumber \eeq
\noindent  where $R \equiv R(r, \theta)$, $P\equiv P(\theta)$ and $Q \equiv Q(r)$ are functions of $\cos \theta$ and $r$, containing the parameters   $m,a, l,$ and $\Lambda$ which are, respectively, mass,  a rotation parameter, a NUT parameter in a de Sitter or anti-de Sitter background, and the cosmological constant:
{\small
\begin{eqnarray}
R^2 &=& r^2 + (l + a\cos \theta)^2, \\ 
P &=& \sin^2 \theta( 1 + (3l+a \cos \theta)(l+a \cos \theta)\Lambda/3), \label{Fns}\\
Q &=& (a^2 - l^2 ) - 2mr + r^2 - \Lambda[(3l^2+a^2)r^2+r^4]/3.
\end{eqnarray}}
\noindent The location of the event horizons are given by the roots of $Q(r)$. 

Following \cite{GANG}, we define the vectors 
\small \beq \begin{aligned} & t^0 = \frac{\sqrt{Q}}{R} \left[ dt - \left(a\sin^2 \theta + 4 l \sin^2 \frac{\theta}{2}\right) d\phi \right],~t^1 = \frac{R}{\sqrt{Q}}dr, &  \\
&t^2 = \frac{\sqrt{P}}{R}\left[ a dt - \left(r^2 +(a+l)^2\right) d\phi \right],~t^3 = \frac{R}{\sqrt{P}} \sin \theta d\theta , \end{aligned} \eeq
\normalsize
and the null frame we will work in is then
\small
\begin{eqnarray}
\begin{aligned} 
& \hat{\ell} = e^{U} \ell = \frac{e^{U}(t^0 - t^1)}{\sqrt{2}},~~ \hat{n} = e^{U} n = \frac{e^{U}(t^0+t^1)}{\sqrt{2}}, & \\ 
&\hat{m} = e^{U} m = \frac{e^{U}(t^2-i t^3)}{\sqrt{2}},~~\bar{\hat{m}} = e^{U} \bar{m}=  \frac{e^{U}(t^2+i t^3)}{\sqrt{2}}.& \end{aligned} \label{tetrad} \eeq
\normalsize
\noindent We assume that the Lorentz frame transformations have been entirely fixed in order to express the SPIs $J_4$ and $K_4$ in a concise form with respect to the Cartan invariants.

Since the metric is of Petrov type {\bf D} we can use $\Psi_2^2$ and its complex conjugate to construct the zeroth order conformally covariant SPIs \cite{AbdLak2015, GANG}:

\beq \hat{I}_1 +i \hat{I}_2 =48 \hat{\Psi}_2^2 = 48 e^{-4U} \Psi_2^2, \eeq
\noindent where 
\beq & \Psi_{2}= -( m + i L) \left( \frac{1}{r +i(l + a \cos \theta)} \right)^3,& \nonumber \eeq
\noindent and the constant, $L$, is defined as
\beq &L = l \left( 1+ \frac13(a^2-l^2)\Lambda \right). & \nonumber \eeq
\noindent Therefore, the simplest zeroth order real valued conformal invariant is 
\beq I_1' = -2 i [\ln \hat{\Psi}_2 - \ln \bar{\hat{\Psi}}_2]. \nonumber \eeq
\noindent At first order, denoting the SPIs $$\hat{J}_{+} = \frac{3 \cdot 2^7}{5} \hat{J}_4, ~ \hat{J}_{-} = \frac{\hat{K}_4}{3\cdot 2^6}, \nonumber $$ \noindent and  defining the vector, $$\hat{v}_a = -\hat{\pi} \hat{m}_a + \hat{\tau} \bar{\hat{m}}_a - \hat{\rho} \hat{n}_a + \hat{\mu} \hat{\ell}_a, \nonumber $$  
\noindent the conformally covariant SPIs $J_{\pm}$ may be expressed in terms of the gradient of $\hat{\Psi}_2$ and ${\bf \hat{v}}$:

{\small
\beq J_{\pm} &=& - 6{\cal R}[ \pm \bar{\hat{\Psi}}_2^2 \hat{\Psi}_2^2 |{\bf \hat{v}}|^2 +    \hat{\Psi}_2^4 |{\bf \hat{v}}|^2] \mp  2 \bar{\hat{\Psi}}_2 \hat{\Psi}_2 \hat{\nabla} \hat{\Psi}_2. \hat{\nabla} \bar{\hat{\Psi}}_2 \nonumber \\ 
&&  + {\cal R}[ \pm \hat{\Psi}_2^2 \hat{\nabla} \bar{\hat{\Psi}}_2.\hat{\nabla} \bar{\hat{\Psi}}_2  - 3 \Psi_2^2 \hat{\nabla} \bar{\hat{\Psi}}_2. \hat{\nabla} \hat{\Psi}_2] \nonumber \\ 
&& + 6 {\cal R} [ \pm \hat{\Psi}_2^3 {\bf \hat{v}}. \hat{\nabla} \hat{\Psi}_2 + \hat{\Psi}_2 \bar{\hat{\Psi}}_2^2 {\bf \hat{v}}. \hat{\nabla} \hat{\Psi}_2]. \nonumber \eeq}

%
 
\noindent  Since the SPIs $J_{\pm}$ transform covariantly under a conformal transformation, we can express the invariants relative to the original coframe for the Kerr-NUT-(Anti)-de Sitter metric, where the spin coefficients satisfy $\pi = \tau$ and $\mu = \rho$, implying that the coframe is an invariant coframe in the context of the Cartan-Karlhede algorithm. Applying the Bianchi identities, the SPIs may be expressed as

{\small
\begin{eqnarray}
e^{10U} \hat{J}_{\pm} = J_{\pm}  &=& -[  \bar{\Psi}_2^4 (\bar{\rho}^2 - \bar{\tau}^2) + \Psi_2^4 ( \rho^2 - \tau^2) \nonumber  \\ && \mp 5 \bar{\Psi}_2^2 \Psi_2^2 ( \bar{\rho}^2 - \bar{\tau}^2 + \rho^2 - \tau^2 ) \label{KNAJ4}  \\ && \pm 12 \bar{\Psi}_2^2 \Psi_2^2 ( \bar{\rho} \rho + \bar{\tau} \tau)], \nonumber
\end{eqnarray} }

\noindent where $\rho$ and $\tau$ are:

\beq \begin{aligned} \rho =  \frac{1}{\sqrt{2}}  \frac{\sqrt{Q} [r - i(a\cos\theta +
    l)]}{R^3 }, \\ \tau = \frac{1}{\sqrt{2}} \frac{a\sqrt{P} [r - i(a\cos\theta + l)]}{R^3 }. \end{aligned} \eeq

\noindent While we have chosen to work with a coframe for which the spin-coefficients and Weyl tensor take a particular form, due to the SPIs' insensitivity to Lorentz frame transformations, the results presented in this section will be valid relative to any coframe.


Since the Kerr-NUT-(Anti)-de Sitter solution has cohomogeneity $n=2$ \cite{GANG}, we may take the conformal invariants: $$I_1' \text{ and }J_4' = J_4 (\Psi_2 \bar{\Psi}_2)^{-\frac52}, \nonumber$$ \noindent to produce the invariant \footnote{The permutation tensor $\delta^{\alpha_1,...,\alpha_n}_{\beta_1,...,\beta_n}$ is $+1$ or $-1$ if $\alpha_1,...,\alpha_n$ is an even or odd permutation of $\beta_1,...,\beta_n$, respectively, and is zero otherwise. }

\begin{eqnarray}
||W||^2 \equiv  \frac{1}{2!}\delta^{a_1 a_2}_{~~~~b_1 b_2} g^{b_1 c_1} g^{b_2 c_2} I_{1;a_1}' J_{4;a_2}' = e^{-2U} |\rho| F , \nonumber
\end{eqnarray} 

\noindent where $F$ is a rational function in terms of $\Psi_2$, its complex conjugate, the spin coefficients and their frame derivatives. The conformally covariant invariant $||W||^2$ detects the event horizons of the Kerr-NUT-(Anti)-de Sitter solution \cite{Page:2015aia} and transforms by being multiplied by a negative power of the conformal factor. 

Taking the sum of the two first order SPIs produces a smaller invariant 
\beq J_{{\it ergo}} = J_{+} + J_{-} = -4 {\cal R} [  \Psi_2^4 ( \rho^2 - \tau^2)]. \label{Jergo} \eeq
\noindent It was noted in \cite{GANG} that the Cartan invariant $\rho^2 - \tau^2$ vanishes on the stationary limit surface. Since the image of the nonvanishing Killing vector field which becomes null on the stationary limit surface is mapped to a nonvanishing conformal Killing vector field which will be null on the image of this surface, we conclude that the SPI, $J_{{\it ergo}}$, will vanish on the conformal stationary limit surface.

\end{section}

\begin{section}{Necessary Conditions for a Dynamical Black Hole to be Conformally Related to a Stationary Black Hole}

While we have considered conformal transformations of stationary black holes to generate dynamical black holes, the problem of determining whether a dynamical black hole is conformally related to a stationary black hole may be hidden by the choice of coordinates. Using the conformal invariants we can introduce a test to determine if a spacetime is conformally related to a spacetime admitting larger maximal dimensional orbits of the isometry group of the local metric and in the case of dynamical black holes introduce a necessary condition for the subcase of dynamical black holes conformally related to a stationary black hole.

To determine if a spacetime is conformally related to another spacetime which admits larger maximal dimensional orbits of the isometry group of the local metric, we will exploit the cohomogeneity, which was shown to be equivalent to the number of functionally independent Cartan invariants produced at the final iteration, $q$, of the Cartan-Karlhede algorithm \cite{GANG}. Setting $p=q-1$, we have the following statement involving the number of functionally independent Cartan invariants of the spacetime, $t_p$, and the number of functionally independent conformal invariants, $\hat{t}_p$.

\begin{prop}
Given a D-dimensional spacetime, $(M, {\bf g})$, if $\hat{t}_p < t_p$ then it is conformal to a spacetime admitting larger maximal dimensional orbits of the isometry group of the local metric.
\end{prop}

Noting that the cohomogeneity of a spacetime is given by the formula $$n = D-r +dim(H_p) = t_p$$ \noindent where $dim(H_p)$ is the dimension of the isotropy group of the spacetime and $r$ is the dimension of the isometry group. If $\hat{t}_p < t_p$ then the spacetime is conformally related to a spacetime, $(M', {\bf g'})$, with $t'_p = \hat{t}_p$ functionally independent Cartan invariants.  This implies that the dimension of the maximal orbits of the isometry group of the new spacetime, $r'-dim(H'_p)$, must be larger than the dimension of the maximal orbit of the isometry group of the original spacetime, $r - dim(H_p)$. 


In the case of dynamical black holes with $\hat{t}_p < t_p$, if the conformally related solution admits a nonvanishing Killing vector field which is stationary and becomes null on some null surface (i.e., contains a Killing horizon) then the norm of the wedge product of the exterior derivatives of the conformal invariants will vanish on a surface. 

\begin{prop}
If a dynamical black hole is conformally related to a stationary black hole, then $\hat{t}_p < t_p$ and the norm of the wedge product of the exterior derivatives of the conformal invariants  will vanish on a surface. 
\end{prop}

\noindent This condition is not sufficient as the SPIs formed by the norm of the wedge products of the exterior derivatives of the conformal invariants may vanish on surfaces that do not correspond to a Killing horizon. 

\begin{subsection}{The Ingoing Kerr-Vaidya Black Hole} 

We examine the Kerr-Vaidya solution \cite{Vaidya1970,IboDor2005,SenTor2014}:
\beq ds^2 &=& -\left(1-\frac{2m(v)r}{R^2} \right) dv^2 + 2 dv dr + R^2 d\theta^2 \nonumber \\ && - \frac{4a m(v) r \sin^2 \theta}{R^2} d\phi dv -2a\sin^2 \theta d\phi dr \label{KerrVaidya} \\ &&+ \frac{(r^2+a^2)^2-a^2 \Delta \sin^2 \theta}{R^2}\sin^2 \theta d\phi^2, \nonumber \eeq
\noindent where $R^2 = r^2 + a^2 \cos^2 \theta$ and $\Delta = r^2+a^2-2m(v)r$.  In general, the cohomogeneity of this solution is $n=3$. By constructing three functionally independent conformal invariants we can determine if this metric is conformally related to a stationary black hole for particular choices of the mass function. 

While the Kerr-Vaidya solution is not of type {\bf D}, it is still of type {\bf II} and so the SPIs $I_1$ and $I_2$ satisfy
\beq I_1+ I_2 = 48 \Psi_2^2  \nonumber \eeq 
\noindent for any frame basis in which the Weyl tensor is of type {\bf II} form\footnote{In particular we employ a similar frame to the null frame used in the Appendix of \cite{IboDor2005}} . Therefore we can construct three functionally independent conformal invariants: 

\beq & I_1' = -2i[\ln \Psi_2 - \ln \bar{\Psi}_2],  (\nabla I_1)' = \frac{\nabla I_1'}{I_1^\frac12},~ J' = \frac{J_4~}{I_1^\frac52}. & \eeq

\noindent Due to the large size of the coordinate expressions of the conformal invariants we will not include them here.

Computing the wedge product of the exterior derivatives of the three invariants yields
{\small
\beq |dI_1' \wedge d(\nabla I_1)' \wedge d(\nabla J)' |^2 = a^3 m F_0 \nonumber	\eeq}

\noindent where $F_0$ is a polynomial in terms of $r$, $\cos \theta$, $m$, $m_{,v}$, $m_{,vv}$ and $m_{,vvv}$. Rewriting this as a polynomial in $r$, setting the coefficients to zero and solving the resulting differential equation for $m(v)$ leads to the solution $m(v) \equiv constant$ or the requirement that $a=0$. Therefore the generic Kerr-Vaidya black hole with $a\neq 0$ and $m_{,v} \neq 0$ is never conformally related to a black hole of lower cohomogeneity. It is possible that the subcase $a=0$ corresponding to the ingoing Vaidya black hole will admit solutions that are conformally related to a stationary black hole.

\end{subsection}

\begin{subsection}{The Ingoing Vaidya Black Hole}
We now consider the ingoing Vaidya solution: 

 \beq ds^2 &=& - \left(1-\frac{2m(v)}{r} \right) dv^2 + 2dv dr \label{IVadiya} \\ && + r^2 (d \theta^2 + \sin^2 \theta d\phi^2), \nonumber \eeq

\noindent where the mass function $m(v)$ is a function of the ingoing null coordinate $v$. In general, this metric has two functionally independent conformal invariants: 

{\small \beq 
J' &=& \frac{J_4}{I_1^\frac52} = \frac{2m_{,v} r^2}{m^2} +2 - \frac{r}{m}, \\
(\nabla J)' &=& \frac{|\nabla J'|^2}{I_1^\frac12} = \frac{2r^2(4m_{,v}r-m)}{m^6}\bigg( (8m_{,v}^2-4 m m_{,vv}) r^3 \nonumber \\
&& -6 m_{,v} m r^2 + (8m_{,v} m^2 + m^2)r -2m^3\bigg).
\eeq}
 
Assuming $m_{,v} \neq 0$ these invariants are functionally independent as the wedge product of their exterior derivatives is nonzero, 

{\small \beq | d J' \wedge d (\nabla J)' |^2 = m_{,vv} F_1(r, m, m_{,v}) + m_{,vvv} F_2(r, m, m_{,v}), \nonumber \eeq}

\noindent where $F_1$ and $F_2$ are polynomials. In order for the conformal invariants to be functionally dependent, we must impose $m_{,vv} = 0$, and so the mass function is now linear, 
\beq m(v) = M_0 + \mu (v-v_0). \nonumber \eeq

The ingoing Vaidya black hole solution has two functionally independent Cartan invariants at zeroth and first order. Since there is at most one functionally independent conformal invariant, we conclude the ingoing Vaidya metric is conformal to a spacetime admitting an additional Killing vector field. To determine when this Killing vector field is potentially null we examine when $J'$ vanishes. The conformal invariant $J'$  vanishes on the surfaces:

\beq r = \frac{m(1 \pm \sqrt{1- 16 m_{,v}})}{4m_{,v}}  =  \frac{m(1 \pm \sqrt{1- 16 \mu})}{4\mu}. \eeq

\noindent Therefore, this can only occur when $\mu \leq 2^{-4}$. This result agrees with \cite{NA2017}, where it was shown that the linear ingoing Vaidya black hole is conformally related to a spherically symmetric static black hole when $\mu \leq 2^{-4}$.
 
\end{subsection}

\end{section}

\begin{section}{Higher Dimensions}

In higher dimensions the invariants $I_1,~ I_3,~ I_{3a}$ and $I_5$ transform in the following manner: 

\small 
\begin{eqnarray}
\tilde{I}_1 &=& e^{-4U} I_1, \label{I1transf} \end{eqnarray}
\begin{eqnarray}\tilde{I}_3 &=& e^{-6U} [ I_3 - (2 I_{1,a} +8 W^e_{~ae} - 8 W_{a~e}^{~e} ) U^{,a} + 8 U_{,a}U^{,a} I_1 \nonumber \\
&&+ (4D-8) U^{,a}U_{,e} C_{abcd} C^{ebcd}], \label{I3transf}\end{eqnarray}
\begin{eqnarray}\tilde{I}_{3a} &=& e^{-6U} [ I_{3a} - ( 2 W^e_{~ae} + I_{1,a} - 2 W_{a~e}^{~e} + 4 \bar{W}^e_{~ea} ) U^{,a}   \nonumber \\
&& + 4 U_{,a}U^{,a} I_1 + (D-1)U^{,a}U_{,e} C_{abcd} C^{ebcd}], \label{I3atransf} \end{eqnarray}
\begin{eqnarray} \tilde{I}_5 &=& e^{-10U} [ I_3 - 8 I_{1,a} U^{,a} + 16 U_{,a}U^{,a} I_1], \label{I5transf}
\end{eqnarray}

\noindent where $W^a_{~bc}$ and $\bar{W}^a_{~bc}$ are defined in equations \eqref{Weqn} and \eqref{Wbeqn}. With some modification to the coefficients of $I_1 I_3,~ I_1 I_{3a}$, and $I_5$ we can produce a higher dimensional analogue to $J_4$: 
\begin{equation}
J_D \equiv  2(D-1) I_1 I_3  - 8(D-2) I_1 I_{3a} +  (D-3)I_5, \label{Jn}
\end{equation} 
\noindent which will vanish on the event horizon of any black hole that is conformal to the $D$ dimensional static spherically symmetric metric \cite{McNuttPage}.

For the conformally transformed metric $\hat{g}_{ab} = e^{2U}g_{ab}$ of a generic nonvacuum metric $g_{ab}$, which is not conformal to a static spherically symmetric metric or a Ricci flat metric, the invariant $\hat{J}_D$ of the metric $\hat{g}_{ab}$ is related to the invariant $J_D$ for the metric $g_{ab}$ by
\begin{equation}
 \hat{J}_D = e^{-10U}(J_D + A^a U_{;a} )
 \label{Jntransform}\;,
\end{equation}
where
\beq \begin{aligned}
 &A_a = -4(D-3) I_{1,a} -  16(W^e_{~ae} -W_{a~e}^{~e}) & \\ &~~~~~~~~+  32(D-2)\bar{W}^e_{~ea}. &  
\end{aligned}  \label{A}\eeq

\noindent Due to the structure of the Weyl tensor in higher dimensions \cite{HDWeyl} the vector $A^a$ will not vanish for all higher dimensional spacetimes. For example, the vector $A^a$ will be nonzero for a generic five dimensional (5D) nonvacuum spacetime of Weyl type {\bf D}. In \cite{McNuttPage} it was shown that $A^a$ will vanish for other stationary spacetimes, such as the Kerr-NUT-Anti-de Sitter black hole solution \cite{generalkerr}, the rotating black ring solution \cite{RBR} and the supersymmetric black ring solution \cite{SBR}. 

The Kerr-NUT-Anti-de Sitter spacetime is of type {\bf D}, and both of the black ring spacetimes are of algebraic type ${\bf I}_i$ according to the alignment classification \cite{HDWeyl,brwands, CHDG}. For any vacuum spacetime, the vanishing of $A^a$ is guaranteed by the differential Bianchi identities \cite{McNuttPage}, and so $A^a$ will vanish for the Kerr-NUT-Anti-de Sitter and the rotating black ring spacetimes. The vanishing of $A^a$ for the supersymmetric black ring metric is notable since it is not Ricci flat. This suggests that the vanishing of $A^a$ is dependent on the form of the Weyl tensor in higher dimensions, and that the differential Bianchi identities for the vacuum spacetimes may give insight into the class of nonvacuum spacetimes for which $J_D$ is conformally covariant.


Unfortunately, these spacetimes have cohomogeneity $n\geq 1 $ \cite{Forget,rbr} and the invariant $J_D$ will no longer detect the horizon. While we would like to construct additional conformally covariant invariants, there are two complications arising from the difference in dimension. First, the Hodge dual no longer maps bivectors to bivectors, and so the Hodge dual of the Weyl tensor is no longer a rank four tensor. Secondly, the structure of the Weyl tensor in higher dimensions no longer ensures that a Weyl-like tensor will yield a conformally covariant first order invariant, therefore for spacetimes where $J_D$ is conformally covariant, a Weyl-like tensor may be produced from the square of the Weyl tensor that does not yield a conformally covariant invariant. 

However, for any spacetime where $J_D$ is conformally covariant, it is possible to generate higher order invariants by combining  $I_1$ and $J_D$ to produce the conformal invariant, $J'_D$, and calculating the norm of its gradient. The differential Bianchi identities and the Ricci identities ensure that the norm of the gradient of $J_D'$ will be functionally independent from $J_D$. Repeating this process we can produce $n$ functionally independent invariants. For example, the cohomogeneity of the 5D Kerr-NUT-Anti-de Sitter metric is $n = \lfloor 5/2 \rfloor=2$. Therefore, the conformal invariants,   
$$J'_D \equiv \frac{J_{D}}{I_1^{\frac52}},~~ K'_D\equiv  \frac{|(\nabla J_{D})'|}{\sqrt{I_1}},$$
\noindent can be used to compute the norm of the wedge product of their exterior derivatives:  
\begin{eqnarray}
||\tilde{W}||^2 \equiv   \frac{1}{2!}\delta^{a_1 a_2}_{~~~~b_1 b_2} g^{b_1 c_1} g^{b_2 c_2} J_{D;a_1}' K_{D;a_2}' .\nonumber
\end{eqnarray} 
\noindent The invariant $||\tilde{W}||^2$ will be nonzero, except on the horizon where it will vanish and it transforms in a covariant manner under a conformal transformation, implying that this invariant will detect the event horizon of any dynamical black hole conformally related to the 5D Kerr-NUT-Anti-de Sitter black hole. 

\end{section}

\begin{section}{Conclusions}

Motivated by the existence of dynamical black holes conformally related to the Schwarzschild metric \cite{SD2005,MMZ2016,Faraoni2009}, we have considered dynamical black holes conformally related to stationary black hole solutions of higher cohomogeneity. In particular, we have examined a generalization of the rotating cosmological black holes described by the Thakurta metric \cite{Thakurta} by applying a conformal transformation to the Kerr-NUT-(Anti)-de Sitter solution. A dynamical black hole generated in this manner has the unusual property that the event horizon could be identified by locating the Killing horizon of the original stationary black hole, {assuming} the conformal factor was known.   

This implies that the event horizon of the dynamical conformal black hole must be a quasi-local surface. Using the fact that the SPI $J_D$ transforms covariantly under a conformal transformation for a larger class of spacetimes than originally intended \cite{McNuttPage}, we have introduced a procedure to generate additional conformally covariant invariants for these spacetimes. This allows for an invariant that can detect the event horizon in a similar manner to stationary black holes \cite{Page:2015aia}, without knowledge of the conformal factor.

In 4D, the trace free condition and algebraic Bianchi identity ensure that any conformally covariant Weyl-like tensor will yield a conformally covariant first order invariant similar to $J_4$. This dimensionally dependent property allows for new functionally independent conformally covariant invariants of higher degree and order to be generated. With a functionally independent set of conformally covariant invariants we can construct a conformally covariant invariant that detects the horizon of any black hole conformally related to a stationary black hole of cohomogeneity $n \leq 3$ \cite{Page:2015aia}. In addition for the dynamical black holes conformally related to the Kerr-NUT-(Anti)-de Sitter solutions, we have shown that the image of the stationary limit surface, the conformal stationary limit surface, can be detected by a conformally covariant SPI.

We have also presented necessary conditions to determine if a $D$-dimensional dynamical black hole is not conformally related to a stationary black hole. The first condition relies on the difference between the number of functionally independent Cartan invariants and conformal invariants for a given spacetime to determine necessary and sufficient conditions for a spacetime to be conformally related to a spacetime of lower cohomogeneity. The second condition relies on the vanishing of the norm of the exterior derivatives of the conformal invariants which is a necessary condition for a stationary horizon.
While we have demonstrated the necessary conditions for the 4D ingoing Kerr-Vaidya and Vaidya black hole solutions, these conditions are applicable in higher dimensions, assuming conformal invariants can be constructed.

In higher dimensions, it is no longer possible to ensure $J_D$ or any invariant constructed in a similar manner will be conformally covariant. However, if a given spacetime has at least two conformally covariant invariants (for example, all nontrivial vacuum spacetimes), then it is still possible to construct new invariants of higher order that are functionally independent and can be combined to form a conformally covariant invariant that detects the stationary horizon. While we have only discussed this construction for the 5D Kerr-NUT-Anti-de Sitter solution, we have verified that this approach is applicable to the 5D rotating black ring and the supersymmetric black ring solutions. 

Due to the structure of the Weyl tensor in higher dimensions, it is of interest to determine if there are other conformally covariant invariants of order $p \geq 1$ besides $J_D$ and its derived invariants. In the case of stationary black holes, the Cartan invariants have been shown to detect the event horizon \cite{GANG,rbr}, thereby providing an alternative set of invariants which are easier to compute than the related SPIs. We believe that by exploring the relationship between $J_D$ and the Cartan invariants, additional conformally covariant invariants may be produced and this will give insight into the equivalence of spacetimes under the conformal group \cite{Skea97, KoutrasSkea98, Lang93}. 

As a final point, the dynamical black holes conformally related to stationary black holes present an opportunity to investigate the geometric horizon conjectures \cite{ADA}; since the event horizon is detected by a SPI, we would like to determine if the Weyl tensor becomes more algebraically special on this surface \cite{HDWeyl}. This property can be determined in an invariant manner using the discrimiant SPIs as necessary conditions \cite{CHDG} or by applying the Cartan-Karlhede algorithm to identify the appropriate coframe \cite{Stephani:2003tm,Forget}.

\end{section}
\begin{section}*{Acknowledgements}  
This work was supported through the Research Council of Norway, Toppforsk grant no. 250367: Pseudo-
Riemannian Geometry and Polynomial Curvature Invariants: Classification, Characterisation and Applications (D.M.). We also would like to thank Malcolm MacCallum, Don Page and Alan Coley for discussions in the early stages of this work. 
\end{section}
%
%

\end{document}